\documentclass{appolx}
\usepackage{epsfig}
\begin{document}
\title{The spin-1/2 Ising model on the bow-tie lattice \\ 
as an exactly soluble free-fermion model\thanks{Presented at CSMAG'07 Conference, 
Ko\v{s}ice, 9-12 July 2007}}
\author{J.~STRE\v{C}KA and L.~\v{C}ANOV\'A
\address{Department of Theoretical Physics and Astrophysics, Faculty of Science, \\ 
P. J. \v{S}af\'{a}rik University, Park Angelinum 9, 040 01 Ko\v{s}ice, Slovak Republic}}
\maketitle
\begin{abstract}
The spin-1/2 Ising model on the bow-tie lattice is exactly solved by establishing a precise 
mapping relationship with its corresponding free-fermion eight-vertex model. Ground-state 
and finite-temperature phase diagrams are obtained for the anisotropic bow-tie lattice 
with three different exchange interactions along three different spatial directions. 
\newline
\end{abstract}

\section{Introduction}
The planar Ising models represent a prominent class of exactly soluble lattice-statistical models, 
which exhibit remarkable phase transitions \cite{baxt89}. Despite the universality of their critical behaviour, it is still of great research interest to analyze accurately the critical behaviour
of a specific lattice with particular interactions. As a matter of fact, the competition between particular interactions might for instance lead to an existence of reentrant phase transitions, 
which have been precisely confirmed in the spin-1/2 Ising model on the anisotropic union 
jack (centered square) lattice \cite{chik87}. In the present work, we shall provide 
the exact solution for the spin-1/2 Ising model on the anisotropic bow-tie lattice.

\section{Model and its exact solution}
Consider the spin-1/2 Ising model on the bow-tie lattice schematically shown in Fig.~1. 
The model under investigation is given by Hamiltonian 
\begin{eqnarray}
{\cal H} = - J_1 \sum_{(i,j)} \sigma_i \mu_j - J_2 \sum_{(k,l)} \sigma_k \mu_l 
           - J_3 \sum_{(m,n)} \sigma_m \sigma_n,
\label{1}	
\end{eqnarray}
where the Ising spins $\sigma_i = \pm 1/2$ and $\mu_j = \pm 1/2$ are used to distinguish 
two kinds of inequivalent lattice sites depicted in Fig.~1 as full and empty circles, respectively, 
\begin{figure}[t]
     \begin{center}
       \includegraphics[width = 0.8\textwidth]{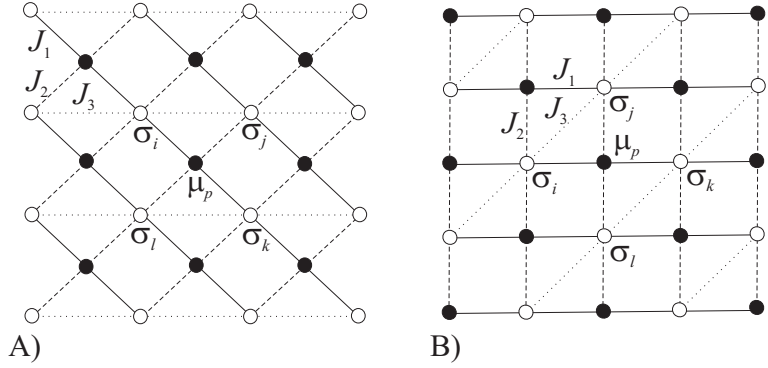}
       \caption{\small A) The magnetic structure of the spin-1/2 Ising model on the bow-tie lattice. 
        The full and empty circles illustrate lattice positions of two inequivalent $\sigma_i$ 
        and $\mu_j$ spin sites, respectively. Solid, dashed and dotted lines schematically 
        represent pairwise spin-spin interactions $J_1$, $J_2$ and $J_3$ along three different 
        spatial directions. B) The bow-tie lattice and its relation with the square lattice.}
       \label{fig1}
     \end{center}
\end{figure}
while the parameters $J_i$ ($i=1-3$) denote pairwise spin-spin interactions along three different 
spatial directions. For further convenience, the total Hamiltonian (\ref{1}) can be rewritten as 
a sum of plaquette Hamiltonians, ${\cal H} = \sum_{p=1}^N {\cal H}_p$, where each plaquette Hamiltonian 
\begin{eqnarray}
{\cal H}_p = - J_1 \mu_p (\sigma_i + \sigma_k) - J_2 \mu_p (\sigma_j + \sigma_l) 
             - J_3 \left(\sigma_i \sigma_j + \sigma_k \sigma_l \right)/2,
\label{2}	
\end{eqnarray}
involves all interactions terms of one square plaquette with the Ising spin $\mu_p $ in its 
centre (the factor 1/2 in the last expression avoids a double counting of the interaction 
$J_3$). In the consequence of that, the partition function of the spin-1/2 Ising model on 
the bow-tie lattice can be partially factorized 
\begin{eqnarray}
{\cal Z} = \sum_{ \{\sigma_i \}} \prod_{p=1}^{N} \sum_{\mu_p = \pm 1/2} \! \! \! \! \!
\exp(-\beta {\cal H}_p) = \sum_{ \{\sigma_i \}} \prod_{p=1}^{N} 
\omega_p (\sigma_i, \sigma_j, \sigma_k, \sigma_l).
\label{3}	
\end{eqnarray}
Above, $\beta = 1/(k_{\rm B} T)$, $k_{\rm B}$ labels the Boltzmann's constant, $T$ is absolute temperature and the Boltzmann's factor $\omega (\sigma_i, \sigma_j, \sigma_k, \sigma_l)$ is given by 
\begin{eqnarray}
\omega (a, b, c, d) = 2 \exp \left[\frac{\beta J_3}{2} \left(ab + cd \right) \right] 
\cosh \left[\frac{\beta J_1}{2} \left(a + c \right) + \frac{\beta J_2}{2} \left(b + d \right) \right].
\label{4}	
\end{eqnarray}
Due to the spin reversal symmetry, there are at best eight different Boltzmann's weights 
to be obtained from Eq.~(\ref{4}) by substituting sixteen allowed spin configurations 
of four corner spins of each square plaquette
\begin{eqnarray}
\omega_1 (+, +, +, +) \! \! &=& \! \! 2 \exp \left(\frac{\beta J_3}{4} \right) 
\cosh \left[\frac{\beta}{2} \left(J_1 + J_2 \right) \right], \nonumber \\
\omega_2 (+, -, +, -) \! \! &=& \! \! 2 \exp \left(-\frac{\beta J_3}{4} \right) 
\cosh \left[\frac{\beta}{2} \left(J_1 - J_2 \right) \right], \nonumber \\
\omega_3 (+, -, -, +) \! \! &=& \! \! 2 \exp \left(-\frac{\beta J_3}{4} \right), \quad
\omega_4 (+, +, -, -) = 2 \exp \left(\frac{\beta J_3}{4} \right), \nonumber \\
\omega_5 (+, -, +, +) \! \! &=& \! \! \omega_6 (+, +, +, -) 
= 2 \cosh \left(\frac{\beta J_1}{2} \right), \nonumber \\
\omega_7 (+, +, -, +) \! \! &=& \! \! \omega_8 (-, +, +, +) 
= 2 \cosh \left(\frac{\beta J_2}{2} \right).
\label{5}	
\end{eqnarray}
In this respect, the Boltzmann's weights (\ref{5}) can readily be regarded as the effective Boltzmann's weights of the corresponding eight-vertex model on a square lattice. It can be easily proved, moreover, that the Boltzmann's weights (\ref{5}) satisfy the free-fermion condition ($\omega_1 \omega_2 + \omega_3 \omega_4 = \omega_5 \omega_6 + \omega_7 \omega_8$) and thus, the spin-1/2 Ising model on the bow-tie lattice is effectively mapped to the free-fermion eight-vertex model solved by Fan and Wu \cite{fan70}. The critical condition of the free-fermion eight-vertex model $\omega_1  + \omega_2 + \omega_3 + \omega_4 = 2 {\rm max} \{\omega_1, \omega_2, \omega_3, \omega_4 \}$ then determines 
the criticality of the spin-1/2 Ising model on the bow-tie lattice provided that the Boltzmann's weights are chosen according to Eq.~(\ref{5}). In our case, the largest Boltzmann's weight is 
either $\omega_1$ or $\omega_2$. Accordingly, if $\omega_1 > \omega_2$ then the critical condition 
reads
\begin{eqnarray}
\exp \left( \frac{\beta_{\rm C} J_3}{2} \right) =
\frac{\cosh \left[\frac{\beta_{\rm C}}{2} \left(J_1 - J_2 \right) \right] + 1}
     {\cosh \left[\frac{\beta_{\rm C}}{2} \left(J_1 + J_2 \right) \right] - 1},      
\label{7}	
\end{eqnarray} 
otherwise
\begin{eqnarray}
\exp \left( \frac{\beta_{\rm C} J_3}{2} \right) =
\frac{\cosh \left[\frac{\beta_{\rm C}}{2} \left(J_1 - J_2 \right) \right] - 1}
     {\cosh \left[\frac{\beta_{\rm C}}{2} \left(J_1 + J_2 \right) \right] + 1},
\label{8}	
\end{eqnarray}
where $\beta_{\rm C} = 1/(k_{\rm B} T_{\rm C})$ and $T_{\rm C}$ is the critical temperature.

\section{Results and discussions}
\begin{figure}[t]
     \begin{center}
       \includegraphics[width = 0.99\textwidth]{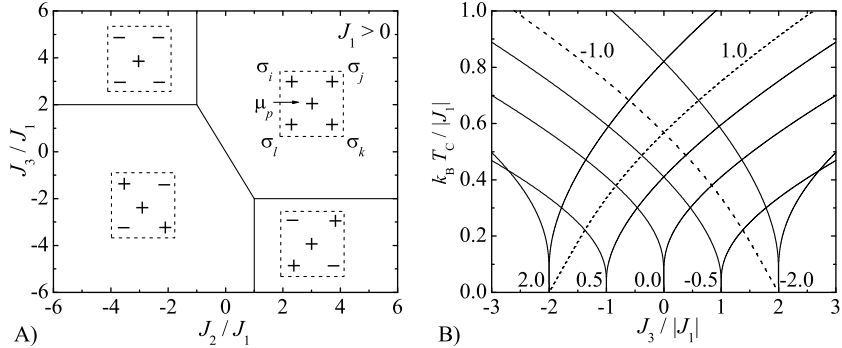}
       \vspace{-0.6cm}
       \caption{\small A) Ground-state phase diagram in the $J_2-J_3$ plane for $J_1>0$. 
       Squares show a typical spin configuration of the basic unit cell of bow-tie 
       lattice within each sector of the phase diagram. The phase diagram for $J_1<0$ case 
       can be obtained by making mirror image of Fig.~\ref{fig2}A) with respect to $J_2 = 0$ axis.
       B) Critical temperature as a function of $J_3/|J_1|$ for several values of 
       the ratio $J_2/|J_1|$.}
       \label{fig2}
     \end{center}
\end{figure}

The most interesting numerical results for the spin-1/2 Ising model on the bow-tie lattice
are displayed in Fig.~\ref{fig2}. Fig.~\ref{fig2}A) illustrates the ground-state phase diagram
with all possible spin configurations that might appear in the model under investigation. 
Furthermore, Fig.~\ref{fig2}B) shows variations of the critical temperature with the ratio 
$J_3/|J_1|$ at several fixed values of $J_2/|J_1|$. It is quite obvious from this figure 
that the finite-temperature phase diagram consists of two wings of critical lines (each 
corresponds to one ordered phase), which merge together with infinite gradient at the 
$T_{\rm C} = 0$ axis whenever $|J_2| \neq |J_1|$. Owing to this fact, the spin-1/2 Ising model 
on the anisotropic bow-tie lattice cannot exhibit similar reentrant phase transitions 
as does the analogous spin-1/2 Ising model on the union jack lattice with a high probability 
because of its over-frustration \cite{chik87}. It is noteworthy that single critical lines, 
which are shown in Fig.~\ref{fig2}B) as broken lines, represent the critical lines for 
two special cases with $|J_2| =|J_1|$. Under this condition, the competing interactions 
give rise to the disordered phase, which appears in the whole parameter space $|J_3|>2|J_1|$ 
due to a strong geometric frustration inherent in the topology of bow-tie lattice.

\end{document}